\documentclass[letterpaper,12pt]{article}

\usepackage{feynmf}

\usepackage{fullpage}

\usepackage[figuresright]{rotating}

\usepackage{amsfonts}

\usepackage{amssymb}

\def\sectionfont{\bfseries\boldmath\rightskip2pc plus1fill{}}

\renewcommand{\baselinestretch}{1.05}

\begin{document}

\begin{fmffile}{F-fmfhp}

\pagestyle{empty}

\begin{flushright}
hep-ph/0211128 \\
MIFP-02-06 \\
ACT-02-09 \\
November, 2002 \\
\end{flushright}

\vspace*{2mm}
\begin{center}
{\Large {\bf F-ENOMENOLOGY}}

\vspace*{8mm}
{\large {\bf D.V. Nanopoulos}}
\vspace*{6mm}

{\it{\small
George~P. and Cynthia~W. Mitchell Institute of Fundamental Physics, \\
Texas A\&M University, \\
College Station, TX 77843-4242, USA \\
\vspace{3mm}
Astroparticle Physics Group, \\
Houston Advanced Research Center (HARC), \\
The Mitchell Campus, \\
The Woodlands, TX 77381, USA \\
\vspace{3mm}
Chair of Theoretical Physics, \\
Academy of Athens, \\
Division of Natural Sciences, \\
28~Panepistimiou Avenue, \\
Athens 10679, Greece \\
\vspace{3mm}
E-mail: dimitri@physics.tamu.edu}}

\vspace*{8mm}

\noindent
\parbox{4.5in}{\footnotesize The advantages of Flipped $SU(5)$ over conventional Supersymmetric $GUT$s, like $SU(5)$, are discussed.
Recent values of $\alpha_s$, $\sin^2\theta_W$, $g_{\mu}\!-\!2$ and the lower limit on the proton lifetime for $p\!\rightarrow\!K^+\bar{\nu}$ {\it point directly} to Flipped $SU(5)$ as the simplest way to avoid potential pitfalls.
It is shown that `F{\it (lipped)}-enomenology' accomodates easily all presently available low-energy data, favoring a rather ``light'' supersymmetric spectrum while yielding the right amount of Cold Dark Matter and a $p\!\rightarrow\!(e^+/\mu^+)\,\pi^0$ lifetime beyond the present experimental limit yet still possibly accessible to a further round of experiments.}

\vspace*{6mm}
\parbox{3.5in}{\hrulefill}
\vspace*{6mm}

\begin{minipage}{4.6in}
{\small
Invited talk given at\,:
\begin{itemize}
\item{The First International Conference on String Phenomenology,\\
\quad Oxford, England, July 6-11, 2002.}
\item{International Workshop on {\it Ne}utrinos and {\it S}ubterranean {\it S}cience,\\
\quad {\it NeSS 2002}, Washington~D.\,C., USA, September 19-21, 2002.}
\end{itemize}}
\end{minipage}

\end{center}

\vspace*{.4cm}

\newpage

\def\thefootnote{\alph{footnote}}
\setcounter{footnote}{0}
\setcounter{page}{1}
\pagestyle{plain}

\section{\sectionfont GRAND UNIFIED THEORIES ($GUT$S)}

The unification of all fundamental interactions observed in Nature is the stuff that physicist's dreams are made of.
The success of Electroweak Unification has opened the way for more grandiose schemes where strong and electroweak interactions unify in the so called Grand Unified Theories ($GUT$s).
There is ample evidence for grand unification, starting from the very particle content of the standard model.
Let us consider the content of the first generation, the next two having identical quantum numbers.
\begin{equation}
\begin{array}{ccccc}
\pmatrix{u\cr d}_L &;& u^c_L &;& d^c_L \\
\vspace{1mm}\\
\pmatrix{\nu_e\cr e}_L &;& e^c_L &;& \fbox{$\nu^c_L$} \\
\end{array}
\quad
\matrix{\textrm{\small{JUST}} \cr \longrightarrow \cr \textrm{\small{RIGHT}}}
\quad
\begin{array}{ccccc}
\pmatrix{d^c_1 \cr d^c_2 \cr d^c_3 \cr e \cr \nu_e}_L &;&
\pmatrix{\pmatrix{u \cr d}_L&\!\!d^c_L&\!e^c_L} &;& \fbox{$\nu^c_L$} \\
\vspace{-.5mm}\\
\bar{\mathbf{5}} &\phantom{;}& \mathbf{10} &\phantom{;}& \fbox{$\mathbf{1}$} \\
\end{array}
\label{eq:smcontent}
\end{equation}

It is most remarkable that the quantum numbers of quarks and leptons, while completely random at the level of the Standard Model, are exactly right so as to fill in the $\mathbf{\bar{5}}$ and $\mathbf{10}$ representation of the minimal possible $GUT$, namely $SU(5)$, as the pioneering work of Georgi and Glashow showed thirty years ago~[1].
For later use I have also included the right-handed neutrino, $\nu^c_L$.
Assuming that such a Grand Unified scheme is realized in Nature, {\it i.e.}~that indeed the Standard Model $SU(3)_c \times SU(2)_L \times U(1)_Y$ is embedded in a simple group $G$, like $SU(5)$, several rather dramatic consequences follow\,:

\begin{enumerate}
\def\theenumi{\roman{enumi}}
\def\theenumii{\alph{enumii}}
\def\labelenumi{(\theenumi)}
\def\labelenumii{(\theenumii)}
\item {\em Charge Quantization}, namely the age-old mystery of why the charges of the proton and positron are exactly the same, is resolved naturally.
The electric charge operator $Q$ is embedded in the simple group $G$ and as such is traceless, $\textrm{Tr}(Q) = 0$, thus providing a corellation between quark and lepton charges.
For example, looking at Eq.\,(\ref{eq:smcontent}), it is seen immediately that $\textrm{Tr}(Q_{\mathbf{\bar{5}}}) = 0$, implying $3Q_{d^c} + Q_e = 0$, and thus that $Q_d = \frac{1}{3}Q_e$, etc.
In a similar mode, the mystery of why quarks and leptons respond equivalently to weak interactions is resolved because this is the way that makes them fit into grand unified representations, which are considered to be the primordial ones.

\item {\em Gauge Coupling Unification} at a very high energy scale M, not far from the Planck Mass ($M_{Pl} \approx 1.2 \times 10^{19}~GeV$), which answers another age-old mystery: the apparent disparity between gauge couplings of strong interactions $(\alpha_3)$ and electroweak interactions $({\alpha_2, \alpha_Y})$ at low-energies such as $M_Z$, the mass of the neutral gauge boson of the $SU(2)_L \times U(1)_Y$ gauge theory.
The gauge coupling unification occurs thanks to the renormalization effects that make the gauge couplings ``run'' and eventually meet at a very high energy scale, due to the fact that the running is logarithmic with energy, and thus needs to cover a large energy range in order for all the couplings to come together~[2].
In a way, the closeness of the grand unified scale $M$ to $M_{Pl}$ prepares the ground for unification with gravity.
If we turn the argument around and assume that it is natural for the grand unified scale $M$ to be close to $M_{Pl}$, then we get naturally the observed pattern of gauge couplings, say at $M_Z$.
More quantitatively, grand unification provides a relation between $\alpha_3(M_Z)$ and $\sin^2\theta_W(M_Z)$ that can be tested experimentally, as will be discussed later.

\item {\em Yukawa Coupling Unification} occurs at the grand unified scale $M$ in a similar way to the gauge coupling unification, {\it i.e.}~Yukawa couplings also ``run'' in quantum field theory~[3,4,5].
A rather dramatic consequence of such a scheme, which was almost immediately confirmed and has remained unscathed for the last twenty five years, is the prediction that~[3,4,5]
\begin{equation}
\lambda_b = \lambda_{\tau} \vert_M  \longrightarrow  m_b \approx 2.5~m_{\tau}\ ,
\label{eq:lambdatau}
\end{equation}

\noindent
if and only if~[4,5] the number of flavors is $N_f = 6$.
It should be stressed that at the time this prediction was made, the primordial nucleosynthesis constraint, coming mainly from the primordial $^4He$ abundance, was $N_f \le 14$~[6].

\item {\em Baryon and Lepton Number Violation} exists due to the very simple fact that quarks and leptons share common $GUT$ representations, and are thus bound to transform directly into each other, thanks to the existence of supermassive grand unified gauge bosons, labeled as $(X_{4/3}, Y_{1/3})$, which belong to the $\mathbf{24}$ adjoint representation of $SU(5)$ and transform as colour triplets and $SU(2)_L$ doublets carrying electric charge $+~4/3$ and $+~1/3$~[4].
Such violations of Baryon and Lepton Number would be fatal, if it was not for the supermassiveness of the mediating gauge bosons $(X , Y)$.
In fact, there is a theorem~[7] proving that in the Standard Model, with it's conventional particle content, Baryon and Lepton Numbers are exact global symmetries, thus ensuring the stability of the proton and the absence of $\mu \rightarrow e \gamma$, in accord with all presently available experimental data.
If though, we move away from the Standard Model, the Baryon and Lepton Number violation floodgates are opened, and one has to check that we still get an experimentally consistent theory.
The answer is that, in principle, not only do we get a consistent theory, but also we have been able to predict several new phenomena which have, for the most part, been verified.
To start with, for the first time, a mechanism to understand the apparent observed matter-antimatter asymmetry in the universe has been proposed, and it is heavily dependent upon Baryon Number violation in $GUT$s~[8].
Furthermore, Lepton Number violation suggests strongly the possibility for neutrino masses, and more specifically, mass differences.
Indeed, the {\it see-saw mechanism} originally suggested by Gell-Mann, Ramond, Slansky and Yanagida~[9], makes use the rather large representation $\mathbf{126}$ of $SO(10)$, and right-handed neutrinos ($\nu^c_L$) as they are naturally present in the $\mathbf{16}$ of $SO(10)$, to give left-handed neutrinos a tiny mass.
\begin{equation}
\textrm{\bf{GRSY:}}
\qquad
\begin{array}{c@{\hspace{2mm}}c}
 & \nu_L \hspace{6mm} \nu^c_L \\
\matrix{\nu_L \cr \nu^c_L} &
\pmatrix{0 & m_D \cr m_D & M} \\
\end{array}
\qquad
\Longrightarrow
\qquad
m_{\nu_L} \approx \frac{m_D^2}{M}
\label{eq:seesaw}
\end{equation}

\noindent
Here, $m_D$ is a ``normal'' Dirac mass and $M$ is close to the unification scale.
One sees that neutrino masses in the (sub) $eV$ range, near the experimental upper bound, are naturally gotten.
Howard Georgi and myself~[10] didn't like the use of such large representations as the $\mathbf{126}$, and proposed an alternate construction where there is an extra singlet $\Phi$ sitting outside the $\mathbf{16}$ of $SO(10)$, and thus leading to a {\it two-step see-saw} mechanism~[10].
\begin{equation}
\textrm{\bf{GN:}}
\quad
\begin{array}{c@{\hspace{2mm}}c}
 & \nu_L \hspace{6mm} \nu^c_L \hspace{6mm} \Phi \\
\matrix{\nu_L \cr \nu^c_L \cr \Phi} &
\pmatrix{0 & m_D & 0 \cr m_D & 0 & M \cr 0 & M & M_U} \\
\end{array}
\quad
\Longrightarrow
\qquad
\begin{array}{rcl}
m_{\nu_L} &\approx& \frac{m_D^2}{M^2 / M_U} \\
\vspace{-1mm}\\
m_{\nu^c_L} &\approx& \frac{M^2}{M_U} \\
\end{array}
\label{eq:twostep}
\end{equation}

\noindent
M indicates, as usual, some Grand Unification scale, and $M_U$ indicates the fundamental scale at which {\it all} interactions in Nature, including gravity, unify, {\it e.g.}~the {\it String Scale}.
A natural first application of such a mechanism was given in the framework of $E_6$ unification~[11], where the $SO(10)$ singlet $\Phi$, is sitting in it's $\mathbf{27}$ representation, which decomposes under $SO(10)$ as $\mathbf{16} + \mathbf{10} + \mathbf{1}$.
The importance of the {\it two-step see-saw} mechanism~[10] will become apparent, through it's modern application, later.
It is heartwarming that the apparent solar $\nu_{e^-}$~deficit as well as the atmospheric $\nu_{\mu^-}$~deficit are most easily explained as neutrino $\nu_e\!-\!\nu_{\mu}$ and $\nu_{\mu}\!-\!\nu_{\tau}$ oscillations respectively~[12].
The production of such oscillations demands that mass differences exist between the neutrino species, indicating~[12] a neutrino mass spectrum in accordance with the predictions of Eq.\,(\ref{eq:twostep})~[10].
As an extra point in support of such neutrino masses, one should mention that the presently favored mechanism for understanding the matter-antimatter asymmetry observed in Nature relies on Leptogenesis, {\it i.e.}~the creation of Lepton Number asymmetry, through the decays of $\nu^c_L$~[13].
This is then recycled into Baryon Number asymmetry by sphalerons at the electroweak scale~[13].
This Leptogenesis $\rightarrow$ Baryogenesis mechanism~[13] entails a $\nu^c_L$ mass pattern again very similar to that given by Eq.\,(\ref{eq:twostep}).
The fact should not escape our notice that the experimentally verified processes of neutrino oscillations imply the existence of right-handed neutrinos $(\nu^c_L)$ to yield the requisite mass terms\footnote{From another point of view, massive particles of only a single handedness cannot exist, as there are Lorentz transformations which will flip this property for $v < c$}, and that these states do not find a home in the minimal $SU(5)~GUT$.
Indeed, one of the best original selling points for $SU(5)$, since there is only room for 15 states within the $\mathbf{\bar{5}}$ and $\mathbf{10}$ representations, was the {\it absence} of $(\nu^c_L)$ from the set of Eq.\,(\ref{eq:smcontent}), and thus the {\it prediction} of massless neutrinos!
This is one of several punch holes which the minimal $SU(5)~GUT$ is subjected to.

At this point, all generic predictions of $GUT$s have been considered and more or less confirmed, at least to the deliberately naive level of sophistication which I have adopted.
Well, all but one, for there remains the issue of\,:
\item {\em Proton Decay}.  Once Baryon-Number conservation has fallen, there is no way to keep protons stable against processes involving the $(X, Y)$ $GUT$ gauge bosons.
For example\,:
\begin{equation}
\fmfframe(3,3)(3,3){
\parbox{30mm}{\begin{fmfgraph*}(30,18)
\fmfleftn{i}{2} \fmfrightn{o}{2}
\fmflabel{$u$}{i1} \fmflabel{$u$}{i2} \fmflabel{$d$}{o1} \fmflabel{$e$}{o2}
\fmf{fermion}{i1,v1} \fmf{fermion}{i2,v1} \fmf{fermion}{o1,v2} \fmf{fermion}{o2,v2}
\fmf{boson,label=$X$}{v1,v2}
\fmfdotn{v}{2}
\end{fmfgraph*}}
$\hspace{5mm} \Longrightarrow \hspace{8mm}
\begin{array}{c}
p \rightarrow e^+\pi^0, \ \tau_p \propto M_X^4\\
\mathbf{d\!=\!6} \end{array}$\ ,}
\label{eq:epizero}
\end{equation}

\noindent
where ``$d\!=\!6$'' reminds us that the effective proton decay operator, after contracting the $X$-boson {\it \`{a} la} Fermi, is a four-fermion interaction of mass dimension $6$, with each field contributing a factor of $\frac{3}{2}$.
Clearly, the proton lifetime, $\tau_p$, is in the $10^{30}$ years range thanks to the superheaviness of the $(X, Y)$ $GUT$ gauge bosons, thus helping one to avoid phobia of self-disintegration.
In principle though, the observation of proton decay remains quite a viable possibility in any $GUT$ framework, and in fact all the other above discussed successes of grand unification more or less imply proton instability at some level.
We will see that not even enormous suppression factors cannot spare the minimal $SU(5)~GUT$ from another punch hole against current experiments.
Having exposed many strongholds of grand unification, it will next be our business to take a closer and more critical look at the available numbers in order to determine how well we really are doing.  Big surprises are in store.
\end{enumerate}

Let us start with gauge coupling unification~[2].
As was mentioned earlier, the gauge couplings ``run'' and can in principle eventually meet at some very high scale $M$.
\begin{eqnarray}
\frac{1}{\alpha_3}-\frac{1}{\alpha_5}&=&\frac{b_3}{2\pi}\,
{\rm \ln}\frac{M}{M_Z}
\nonumber\\
\frac{1}{\alpha_2}-\frac{1}{\alpha_5}&=&\frac{b_2}{2\pi}\,
{\rm \ln}\frac{M}{M_Z}
\label{eq:running}\\
\frac{1}{\alpha_Y}-\frac{1}{\alpha_5}&=&\frac{b_Y}{2\pi}\,
{\rm \ln}\frac{M}{M_Z}
\nonumber
\end{eqnarray}

\noindent
In the above, $\alpha_5$ is the grand unified gauge coupling at the scale M, where $\alpha_3$, $\alpha_2$ and $\alpha_Y$\footnote{While the Standard Model couplings can be taken in the equations of\,(\ref{eq:running}) at their $M_Z$ values, the relations remain valid when this scale is replaced by any value $Q$ between $M_Z$ and $M$.} are supposed to meet, and $b_{i = 1,2,3}$ depend on the well-tabulated content of the Standard Model.
We see also that there are two unknowns, ($\alpha_5$, $M$) and three equations, implying a relation between $\alpha_3$, $\alpha_2$ and $\alpha_Y$, which usually is taken to be $\sin^2\theta_W = \frac{1}{6} + \frac{5~\alpha_{em}(M_Z)}{9~\alpha_3(M_Z)}$.
It turns out that this relation fails if we plug in presently available data for $\alpha_3$, $\alpha_2$ and $\alpha_{em}$, the electromagnetic fine-structure constant.
To put this another way, $\alpha_3$, $\alpha_2$ and $\alpha_Y$ {\it do not} meet at a unique high energy scale $M$, and so much for gauge coupling unification at a single point, such as in the minimal $SU(5)~GUT$.
One may still take take heart for the premise of unification due to the {\it near} convergence which is seen, but for now hope is followed by disappointment, as the ballpark $GUT$ scale $M \approx O(10^{15}~GeV)$ which emerges implies a proton lifetime $\tau_p \approx O(10^{30} y)$, much shorter than the presently available lower bounds (see below).
{\it Could there be some path leading out of this quagmire?}

\section{\sectionfont SUPERSYMMETRIC ($SU\!SY$) $GUT$S}

These ``grand'' failures of the minimal $SU(5)~GUT$ indicated that we should have taken a different path to unification.
Surprisingly enough, unlike Robert Frost's ``Road not taken'', here it is a road which had already been taken even before the conflicts above were exposed.
Let us recall that there is a ``grand'' energy gap between electroweak unification at some characteristic scale, say $M_Z \approx 91~GeV$ and the grand unified scale $M \approx O(10^{15}~GeV)$.
This huge disparity between the two energy scales immediately creates the cumbersome difficulty known as the {\it gauge hierarchy problem}.
The problem is that higher order quantum correction try to erase the hierarchy ($M \gg M_Z$) by shifting all low scales close to $M$!
What really happens is that the mass of the electroweak Higgs boson, assumed to be $m_h \approx O(100~GeV - 1~TeV)$ after the spontaneous breakdown which it triggers, gets quadratically divergent quantum corrections.
These corrections imply $\delta m_h^2 \approx \alpha_2 M^2$, where the $GUT$ scale $M$ provides a natural cutoff.
This is, of course, catastrophic and should be certainly avoided.
The way out of this serious problem has been provided by a new symmetry in Nature that relates fermions and bosons, called {\it Supersymmetry}, or $SU\!SY$~[14,15].
The idea here is that for every fundamental particle in Nature, there exists a partner which exactly duplicates all of it's quantum numbers, modulo spin, which is shifted plus or minus by $\frac{1}{2}$ unit.
Exact $SU\!SY$ entails equal masses for each particle and it's {\it s}(uper){\it particle} companion, but the defining mark of broken $SU\!SY$ is a divergence of this symmetric mass relation between the partners by an amount proportional to the $SU\!SY$ breaking scale, $\tilde{m}$.
Supersymmetric field theories have much better quantum behavior, {\it i.e.}~much less severe divergences, due to the fact that, all else being equal, quantum loops involving fermions versus bosons have a relative sign difference in accordance with Fermi-Dirac versus Bose-Einstein statistics.
Because of this, many cancellations between related Feynman diagrams do occur.
For the case in hand, exact $SU\!SY~GUT$s will entail the complete absence of quadratic Higgs mass corrections, and the more realistic case of broken $SU\!SY$ predicts $\delta m_h^2 \approx \alpha_2 \tilde{m}^2$.
There are no hitherto observed Standard Model particles which may suitably be interrelated by supersymmetry, and clearly this lack of direct evidence for sparticles must impose a lower bound of $\tilde{m} \geq O(100~GeV)$.
Conversely, the attractive $SU\!SY$ resolution of the gauge hierarchy problem demands that $\tilde{m} \leq O(1~TeV)$, {\it i.e.}~around the electroweak scale.
Amusingly enough, the more than doubling of the Standard Model particle spectrum, through superpartners at the electroweak scale, naturally resolves several of the other previously mentioned problems of the minimal $SU(5)~GUT$.
The very existence of sparticles allows for new types of loops in the renormalization diagrams for the extra particle content to circulate, and thus alters the values of the $\beta$-function coefficients ($b_{1,2,3}$) in Eq.\,(\ref{eq:running}).
This has the effect of naturally postponing grand unification until $M \approx O(10^{16}~GeV)$~[16], which in turn leads to two most beneficial consequences.
Namely, the predicted proton lifetime increases (see Eq.\,(\ref{eq:epizero})) to $\tau_p \approx O(10^{34}~y)$, thus bypassing the experimental lower bound, and the correlation between $\alpha_3(M_Z)$ and $\sin^2\theta_W(M_Z)$ is now satisfied\footnote{In other words, the notion of a single high-energy convergence of the three gauge couplings becomes much more nearly compatible with low-energy phenomenology.} ($\sin^2\theta_W = \frac{1}{5} + \frac{7~\alpha_{em}(M_Z)}{15~\alpha_3(M_Z)}$), {\it grosso-mondo}~[17].
This is not a bad start for a new symmetry!
One may boldly suggest that the ``observed'' gauge coupling unification at $M \approx O(10^{16}~GeV)$ is a sensitive {\it supermatter meter}.
Low-energy supersymmetry is the {\it only} known extension of the Standard Model that is in {\it super}b agreement with all low energy data, most notably that from $LEP$, which has provided the best available electroweak precision measurements.
In addition, $SU\!SY$ has provided very natural, {\it dynamic} means to achieve the electroweak spontaneous breakdown itself, in the form of radiative corrections~[15].
This demanded a rather heavy top quark~[15] as was subsequently discovered experimentally!
The Lightest Supersymmetric Particle (LSP), the neutralino ($\chi^0$) is stable and it has exactly the right quantum numbers and interactions to provide for the astrophysically long-sought cold dark-matter in the Universe~[18], {\it a rather unexpected prediction}!
The new supersymmetric degress of freedom provide extra ways for the proton to decay, such as~[19,20]\,:
\begin{equation}
\fmfframe(3,3)(3,3){
\parbox{40mm}{\begin{fmfgraph*}(40,18)
\fmfleftn{i}{2} \fmfrightn{o}{2}
\fmflabel{$s$}{o1} \fmflabel{$\nu$}{o2}
\fmfforce{0,.15h}{i1} \fmfforce{0,.85h}{i2}
\fmf{fermion,tension=.25,label=$u$,l.side=right}{i1,v1} \fmf{fermion,tension=.25,label=$d$,l.side=left}{i2,v2}
\fmf{plain,label=$\tilde{W}$,l.side=left,l.dist=15}{v1,v2}
\fmffixedy{.7h}{v1,v2}
\fmf{dashes,tension=.4,label=$\tilde{d}$,l.side=right,l.dist=15}{v1,v3}
\fmf{dashes,tension=.4,label=$\tilde{u}$,l.side=left,l.dist=15}{v2,v3}
\fmf{plain,tension=.75,label=$\tilde{H}_{\mathbf{3}}$,l.side=left}{v3,v4}
\fmf{plain,tension=.75,label=$\tilde{H}_{\mathbf{\bar{3}}}$,l.side=left}{v4,v5}
\fmf{fermion}{o1,v5} \fmf{fermion}{o2,v5}
\fmfdot{v1,v2,v3,v5}
\fmfv{decoration.shape=cross}{v4}
\end{fmfgraph*}}
$\hspace{5mm} \Longrightarrow \hspace{8mm}
\begin{array}{c}
p \rightarrow K^+\bar{\nu}, \ \tau_p \propto M_{\tilde{H}_{\mathbf{3}}}^2\\
\mathbf{d\!=\!5} \end{array}$\ ,}
\label{eq:kplusnu}
\end{equation}

\noindent
where squarks ($\tilde{u},\tilde{d}$), winos ($\tilde{W}$) and color triplet higgsinos  ($\tilde{H}_{\mathbf{3}},\tilde{H}_{\mathbf{\bar{3}}}$) participate.
The Baryon and Lepton Number violation occurs through the effective $\tilde{u} \tilde{d} \nu s$ operator, after the superheavy $\tilde{H}_{\mathbf{3}} \tilde{H}_{\mathbf{\bar{3}}}$ contraction, which consists of two scalars ($\tilde{u},\tilde{d}$) and two fermions ($\nu, s$), and is thus an operator of {\it mass dimension} $d\!=\!5$.
As such, one finds $\tau_p \propto M^2_{\tilde{H}_{\mathbf{3}}}$ for the corresponding proton lifetime, which is {\it much shorter} than the one provided by the $d\!=\!6$ operators (see Eq.\,(\ref{eq:epizero})), and a potential source of big trouble~[19].
It is worth pointing out that the origin of $d\!=\!5$ proton decay operators is directly traced to an {\it extraneous} fine-tuning that is necessary in $SU\!SY~SU(5)$, in the following way.
In the Minimal Supersymmetric Standard Model ($M\!S\!S\!M$) one needs two Higgs doublets, $H_{\mathbf{2}}$ and $H_{\mathbf{\bar{2}}}$, in order to provide masses to all quarks and leptons, which get promoted to a $h_{\mathbf{5}} \equiv (H_{\mathbf{2}}, H_{\mathbf{3}})$ and a $h_{\mathbf{\bar{5}}} \equiv (H_{\mathbf{\bar{2}}}, H_{\mathbf{\bar{3}}})$ in $SU\!SY~SU(5)$.
The color triplet Higgs bosons and Higgsinos do mediate proton decay, thus they need to be superheavy, close to the $SU\!SY~GUT$ scale $M \simeq O(10^{16}~GeV)$.
This means that they need to obtain their masses from the $GUT$ spontaneous breakdown, which is easy to achieve since there is a natural $h_{\mathbf{5}} h_{\mathbf{\bar{5}}} \Sigma_{\mathbf{24}}$ coupling, where $\Sigma_{\mathbf{24}}$ denotes the adjoint Higgs responsible for the $GUT$ breaking.
However, the Higgs doublets then become super-heavy, which is a disaster!
Back to another gauge hierarchy problem?
Yes, unless we prevent it with {\it extraneous} fine-tuning, by introducing another term $M h_{\mathbf{5}} h_{\mathbf{\bar{5}}}$, and playing it against the $\Sigma_{\mathbf{24}}$ term so that we fix by hand a ``massless'' $H_{(\mathbf{2},\mathbf{\bar{2}})}$ and a super-massive $H_{(\mathbf{3},\mathbf{\bar{3}})}$.
This is arguably {\it not} an ingenious suggestion, for one then arrives at the notorious {\it Higgs doublet-triplet splitting problem}.
Along with the desired mass term, we also {\it necessarily} receive an accompanying mixing between the Higgsinos, $M H_{\mathbf{3}} H_{\mathbf{\bar{3}}}$ (represented by the cross-$\!\mathbf{\times}$ sign in Eq.\,(\ref{eq:kplusnu})), which then cancels one of the two fermion propagators, leading to a reduced $\tau_p \propto M^{2}_{\tilde{H}_\mathbf{3}}$.
In other words, there exists a deeply-rooted connection between the Higgs double-triplet splitting problem and the $d\!=\!5$ proton decay operators.
One may say that the short dimension five proton lifetime that runs into potential trouble with the presently available lower bounds is  a quantitative proof that Nature {\it abhors} extraneous fine-tuning.
Alas, the minimal $SU\!SY~SU(5)$ has even more acute problems.
One is the the exact value of $\sin^2\theta_W$, which acquires important corrections from threshold effects both at the electroweak and $GUT$ scale.
Precise measurements,
\begin{eqnarray}
\alpha_s (M_Z) &=& 0.1185 \pm 0.002
\nonumber\\
\sin^2\theta_W &=& 0.231173 \pm 0.00016
\label{eq:consts}\\
\alpha_{EM} &=& \frac{1}{127.943 \pm 0.027}\ ,
\nonumber
\end{eqnarray}

\noindent
indicate a conflict in the correlation between $\alpha_s(M_Z)$ and $\sin^2\theta_W$, even in $SU\!SY~SU(5)$.
Another problem is the lifetime of the proton.
Minimal $SU\!SY~SU(5)$ avoids, as was discussed above, the catastrophically rapid $p \to e^+\pi^0$ decay that ``did in'' non-$SU\!SY~SU(5)$; however it predicts $p \to K^+ \bar{\nu}$ at a rate too fast to satisfy the presently available lower limit on the lifetime for this decay,
\begin{equation}
\tau (p \rightarrow K^+\bar{\nu}) \ge 6.7 \times 10^{32} y\ ,
\label{eq:kplusnulim}
\end{equation}

\noindent
at the 90\% confidence limit.
The latter requires the $SU(5)$ color-triplet Higgs $H_{\mathbf{3}}$ particles to weigh $ \ge 7 \times 10^{16}~GeV$, whereas conventional $SU(5)$ unification, respecting Eq.\,(\ref{eq:consts}), would impose the {\it upper limit} of $3.6 \times 10^{15}~GeV$ at the 90\% C.L.~[21].
As Table~1 clearly displays, minimal $SU\!SY~SU(5)$ seems to be failing all important tests of Grand Unification, or to put it more directly, this model seems to be {\it excluded} by present experimental data!
Although Nature does not seem to make great use of {\it minimal} $SU(5)$, {\it flipped} $SU(5)$ does however seem to be a favored model, as the table also indicates.
{\it What then is flipped $SU(5)$, and how does it come into the picture?}
\begin{table}[ht]
\center{Table 1: Comparison between $SU(5)$ and flipped $SU(5)~GUT$ features.}
\vspace{2mm}
\label{tab:1}
\begin{center}
\begin{tabular}{|l|c|c|}\hline
Basic GUT tests&$SU(5)$&Flipped $SU(5)$\\ \hline
$\sin^2\theta_W\Rightarrow\alpha_3(M_Z)$&$\times$&$\surd$\\
Proton decay&$\{p\to \bar\nu K^+\}~\times$&$\{p\to (e^+/\mu^+)\pi^0\}$\\
Doublet-triplet splitting&$\times$&$\surd$\\
Neutrino masses&$\times$&$\surd$\\
Baryogenesis&$\times$&$\surd$\\ \hline
\end{tabular}
\end{center}
\vspace{-0.6cm}
\end{table}

\section{\sectionfont FLIPPED $SU(5)$ UNIFICATION}

Grand Unification is a great idea and it should play an important role in the construction of the Theory of Everything (TOE).
We should find a way to avoid all of the grave problems that $SU\!SY~SU(5)$ is mired in and keep all of the ``good stuff''.
An extreme alternative is to abandon Grand Unification completely and jump directly from the Minimal Supersymmetric Standard Model to the TOE ({\it e.g.}~String\,/$M$-Theory).
In that case, most of the ``goodies'' discussed above look accidental and irrelevant.
For example, Luis Ib\'a\~nez has recently suggested~[22] that the logarithmic unification of the gauge couplings is as {\it fortuitous} as the apparent similarity in the sizes of the sun and moon as viewed from the earth!!!
No Comment.
Another way to proceed is to enlarge $SU(5)$ to $SO(10)$, {\it etc.} with the hope that maybe the problems endemic to $SU(5)$ will disappear.
While this is a noble enterprise, one becomes frightened by the out-of-hand plethora of large representations that one is forced to deal with, such as $\mathbf{45}$, $\mathbf{120}$, and $\mathbf{126}$ in SO(10), $\mathbf{27}$, $\mathbf{78}$, and $\mathbf{351}$ in $E_6$ {\it etc., etc.}
All of these extensions of $SU(5)$ inherit some of it's problems, {\it e.g.}~the Higgs double-triplet splitting, as well as over-constrained and problematic fermion mass relations, even if they may be concocted in such a way to avoid, say the proton-decay limits.
Another problem that all of these models are facing is that if they at some stage try to become string-derived models, then it is difficult, and {\it impossible} at the level of $k\!=\!1$ perturbative string theory, to get all of the needed large representations~[23].
Somehow the aesthetically unappealing use of large representations receives a dramatic ``thumbs down'' signal from String Theory~[23].
So what's left?
In other words, is there a there a ``Grand Unified'' Theory with ``minimal'' types of representations, like the ($\mathbf{5},\bar{\mathbf{5}}$)'s and ($\mathbf{10},\bar{\mathbf{10}}$)'s of $SU(5)$ that encompasses {\it just} the ``Standard'' particles, resolves all the problems mentioned above and is string derivable?
I am aware of only one such theory, {\it flipped $SU(5)$}~[24,25]!
Flipped $SU(5)$ is based on the gauge group $SU(5) \times U(1)$.
In close analogy with $SU(2)_L \times U(1)_Y$, the electric charge operator, $Q_{EM}$ is shared between $SU(5)$ and $U(1)$.
This necessarily leads to the following reshuffling of quarks and leptons in Eq.\,(\ref{eq:smcontent})\,:
\begin{equation}
f_{\mathbf{\bar{5}}} = \pmatrix{u^c_1\cr u^c_2\cr u^c_3\cr e\cr \nu_e}_L \quad ; \quad
F_{\mathbf{10}} = \pmatrix{\pmatrix{u \cr d}_L&\!\!d^c_L&\!\nu^c_L} \quad ; \quad
l_{\mathbf{1}} = e^c_L\ ,
\label{eq:flippedcontent}
\end{equation}

\noindent
and similarly for the next two generations.
One immediately notices the flipping $u^c_L \Leftrightarrow d^c_L$ and $\nu^c_L \Leftrightarrow e^c_L$, thus the moniker {\it Flipped $SU(5)$}!
This innocuous flipping has rather dramatic consequences.
While Eq.\,(\ref{eq:flippedcontent}) is referring to quarks and leptons, the quantum numbers will be identical for any other particles fitting into these representations.
In particular, we see that the $\mathbf{10}$ representation has a neutral component with the quantum numbers of $\nu^c_L$.
We may achieve spontaneous $GUT$ breaking by using a $\mathbf{10}$ representation, along with a $\mathbf{\bar{10}}$ companion, for the superheavy Higgs in $SU\!SY~SU(5) \times U(1)$, where the neutral components develop a large vacuum expectation value ({\it vev}), $\langle\nu^c_H\rangle = \langle\bar{\nu}^c_H\rangle$.
\begin{equation}
H_{\mathbf{10}} = \{ Q_H , d^c_H , \nu^c_H \} \quad;
\quad H_{\mathbf{\bar{10}}} = \{ Q_{\bar{H}} , d^c_{\bar{H}} , \nu^c_{\bar{H}} \}\ ,
\label{eq:higgsvev}
\end{equation}

\noindent
while the electroweak spontaneous breaking occurs through the doublets $H_\mathbf{2}$ and $H_{\mathbf{\bar{2}}}$, exactly as in $SU(5)$ mentioned above.
\begin{equation}
h_{\mathbf{5}} = \{ H_{\mathbf{2}}, H_{\mathbf{3}} \} \quad;
\quad h_{\mathbf{\bar{5}}} = \{ H_{\mathbf{\bar{2}}}, H_{\mathbf{\bar{3}}} \}
\label{eq:gutdoub}
\end{equation}

\noindent
Once more, the close analogy between $SU(2) \times U(1)$ and $SU(5) \times U(1)$ is rather apparent, since the quarks and leptons, as well as {\it all} of the Higgs particles needed ($GUT$ and electroweak) are in identical representations of the {\it minimal possible} $(\mathbf{5}, \mathbf{\bar{5}})$ and $(\mathbf{10}, \mathbf{\bar{10}})$.
The $GUT$ superpotential takes the form\,:
\begin{equation}
W_G = H H h + \bar{H} \bar{H} \bar{h} + F \bar{H} \Phi + \mu h \bar{h}\ ,
\label{eq:gutsup}
\end{equation}

\noindent
where $\Phi$ refers to an overall ${SU(5) \times U(1)}$ singlet, alluded to before (see Eq.\,(\ref{eq:twostep})), and the
supersymmetric ``$\mu - term$'' provides the necessary $h\!-\!\bar{h}$ mixing to disallow problematic electroweak axions, with $\mu$ in the range ${100~GeV \to 1~TeV}$.
Interestingly enough, the remaining terms in the $GUT$ superpotential $W_G$ serve also noble causes, such as\,:
\begin{itemize}
\item {\em Natural Higgs doublet-triplet splitting}~[25].
\end{itemize}
As explained above, the components of the Higgs pentaplets must be split,
\begin{equation}
h_{(\mathbf{5},\mathbf{\bar{5}})} =
\pmatrix{H_{(\mathbf{2},\mathbf{\bar{2}})} \cr H_{(\mathbf{3},\mathbf{\bar{3}})}}
\quad \Rightarrow \quad
\pmatrix{\textrm{\small{ELECTROWEAK SYM. BREAKING}} \cr \textrm{\small{PROTON DECAY}}}\ ,
\label{eq:pentsplit}
\end{equation}

\noindent
because of their very different roles.
The interactions in $W_{G}$,
\begin{eqnarray}
H  H h &\rightarrow& d^c_H \langle\nu^c_H\rangle H_{\mathbf{3}}
\nonumber\\
\bar{H} \bar{H} \bar{h} &\rightarrow& \bar{d}^c_H \langle\bar{\nu}^c_H\rangle H_{\mathbf{\bar{3}}}\ ,
\label{eq:nuvev}
\end{eqnarray}

\noindent
make the triplets heavy, while leaving the doublets light.
This is the {\it missing partner mechanism}~[25]!
Most importantly, this {\it dynamic} doublet-triplet splitting does not need or involve any $GUT$-scale ($O(M)$) mixing between $H_{\mathbf{3}}$ and $H_{\mathbf{\bar{3}}}$, thus leading to a
\begin{itemize}
\item {\em Natural suppression of $d\!=\!5$ proton decay},
\end{itemize}
(see Eq.\,(\ref{eq:kplusnu})) to the order of $(\frac{\mu}{M_{\bar{H}}})^2$~[25]!
Indeed, as discussed above the relation between the doublet-triplet splitting and $d\!=\!5$ proton decay is deeply-rooted; one stone, two birds.

The Yukawa superpotential $W_Y$ takes the form
\begin{equation}
W_Y = \lambda_u F \bar{f} \bar{h} + \lambda_d F F h + \lambda_e \bar{f} h^c h\ ,
\label{eq:yuksup}
\end{equation}

\noindent
as entailed by the ``flipping'' of the assignments of the Standard Model fields to the $SU(5)$ representations.
Such a ``flipping'' brings the $\nu^c$ field into the $F$ representation, providing a source of Dirac neutrino masses, as well as a see-saw type neutrino coupling out of $(F \bar{H} \Phi)$ in $W_G$ (see Eq.\,(\ref{eq:gutsup})).
\begin{eqnarray}
\lambda_u F \bar{f} h &\rightarrow& m_u \nu \nu^c
\label{eq:nudiraca}\\
F \bar{H} \Phi &\rightarrow& \langle\nu^c_H\rangle \nu^c \Phi\ ,
\label{eq:nudiracb}
\end{eqnarray}

\noindent
This leads {\it directly} to a
\begin{itemize}
\item {\em Natural two-step see-saw Mechanism}, {\it \`{a} la} GN~[10],
\end{itemize}
reproducing Eq.\,(\ref{eq:twostep}) {\it exactly}, with $M \approx O(\langle \nu^c_H\rangle)$~[25].
Remarkably, it has been shown that such a ``flipped'' neutrino mass matrix easily accomodates~[26], and even foresaw~[27], all presently available neutrino physics data~[12], while it provides a natural source~[28] of lepton asymmetry that, as discussed previously, is the main ingredient in the presently favoured~[13] explanation of baryogenesis, or matter-antimatter asymmetry, observed in the the universe~[28].

Until now, we have seen that the ``sharing'' of the photon between $SU(5)$ and $U(1)$, which necessarily leads to $u^c_L \Leftrightarrow d^c_L$ and  $\nu^c_L \Leftrightarrow e^c_L$, has provided astonishingly elegant solutions to the grave problems of ``canonical'' $SU(5)$, as summarized in Table~1.
Once more, one is tempted to draw on the close analogy between $SU(2) \times U(1)$ and $SU(5) \times U(1)$ in predicting new particles and interactions.
In the case of $SU(2)_L \times U(1)_Y$, the ``cohabitation'' of the photon in $SU(2)_L$ and $U(1)_Y$ makes natural the appearance of a second neutral gauge boson, the $Z^0$, which mediates a new type of neutral current in addition to $QED$.
Both the coupling strength of the neutral currents and the mass of the $Z^0-$gauge boson ($M_Z$) are functions of $\sin^2\theta_W$ and their discovery, in accordance with theoretical predicitions, heralded the dawn of the Electroweak era.
Similarly, the ``sharing'' of the photon between the $SU(5)$ and $U(1)$ leads one to the $\nu_L^c \Leftrightarrow e_L^c$ ``flipping'', which then necessitates the introduction of a new set of particles, the right-handed neutrinos $\nu_L^c$.
The masses $m_{\nu_L^c}$ will be as given in Eq.\,(\ref{eq:twostep}), depending on $\sin^2\theta_W$ through the value of $M$, and then naturally lead to neutrino oscillations as have since been observed in solar and atmospheric deficits~[12]!
Notice that in the case of $SU(2) \times U(1)$, Nature could have chosen the more elegant ``grand unified'' Georgi-Glashow $SU(2)$ model~[29], avoiding the weak neutral currents.
Ditto for the grand unified Georgi-Glashow $SU(5)$ model~[1] which avoids the right-handed neutrinos, and along with them, neutrino masses and neutrino oscillations (see Table~2).
{\it But She didn't!!!}
\begin{table}[ht]
\center{Table 2: Comparison between unified and grand unified\\$SU(2)$ and $SU(5)$ gauge
groups and their properties.}
\vspace{2mm}
\label{tab:2}
\begin{center}
\begin{tabular}{|l|l|}
\hline
\underline{SU(2)}
&\underline{SU(2)$\times$U(1)}\\
{\small[Georgi-Glashow `72]}
&{\small[Glashow `61, Weinberg `67, Salam `68]}\\
$\bullet$ ``grand" unified
&$\bullet$ unified\\
$\bullet$ $W^\pm,\gamma$: $\gamma$ inside $SU(2)$
&$\bullet$ $W^\pm,Z,\gamma$: $\gamma=\{W^3$ [$SU(2)$], $B$ [$U(1)$]$\}$\\
$\bullet$ Higgs triplet (adjoint)
&$\bullet$ Higgs {\it doublet}, {\it \`a la} quarks, leptons\\
$\bullet$ Neutral currents exist (1973)
&$\bullet$ $SU(3)$ not accounted for;\\
 &\quad grand unification later\\
$\bullet$ Wrong!
&$\bullet$ Right!\\
\hline
\hline
\underline{SU(5)}
&\underline{SU(5)$\times$U(1)}\\
{\small[Georgi-Glashow `74]}
&{\small[Barr `82, Derendinger-Kim-Nanopoulos `84,}\\
 &\quad {\small Antoniadis-Ellis-Hagelin-Nanopoulos `87]}\\
$\bullet$ grand unified
&$\bullet$ unified\\
$\bullet$ $W^\pm,W^3,B,X,Y$
&$\bullet$ $W^\pm,W^3,B,X,Y,\widetilde B$\\
$\quad \gamma$ inside $SU(5)$
&$\quad\gamma$: $(W^3,B)$ [$SU(5)$], $\widetilde B$ [$U(1)$]\\
$\bullet$ Higgs $\mathbf{24}$ (adjoint)
&$\bullet$ Higgs $\mathbf{10}$,$\mathbf{\bar{10}}$ (antisymmetric),\\
 &\quad {\it \`a la} quarks, leptons\\
$\bullet$ $\alpha_3(M_Z)>0.13$;
&$\bullet$ Gravity$^*$ not accounted for;\\
\quad $\tau~(p \to K^+\bar{\nu})$ {\it too short}
&\quad grand unification later\\
$\bullet$ Wrong!
&$\bullet$ Right?\\
\hline
\end{tabular}
\end{center}
\vspace{-5mm}
\center{\small{$^*$ ``Gravity" = supersymmetry breaking, hidden sector gauge groups, string unification.}}
\vspace{-1mm}
\end{table}

\section{\sectionfont FLIPPED PHENOMENOLOGY (F-ENOMENOLOGY)}

As will be shown next, the ``gauge coupling unification'' problem of $SU(5)$ finds a natural resolution in $SU(5) \times U(1)$, once more thanks to the ``cohabitation'' property of the photon.
In flipped $SU(5)$, there is a first unification scale, $M_{32}$, where the $SU(3)$ and $SU(2)$ gauge coupling become equal, given to lowest order by~[30]
\begin{eqnarray}
\frac{1}{\alpha_3}-\frac{1}{\alpha_5}&=&\frac{b_3}{2\pi}\,
{\rm \ln}\frac{M_{32}}{M_Z}\ ,
\nonumber\\
\frac{1}{\alpha_2}-\frac{1}{\alpha_5}&=&\frac{b_2}{2\pi}\,
{\rm \ln}\frac{M_{32}}{M_Z}\ ,
\label{eq:m32run}
\end{eqnarray}

\noindent
where $\alpha_2 = \alpha/\sin^2\theta_W$, $\alpha_3 = \alpha_s(M_Z)$ and the one-loop beta-functions are $b_2 = +1$, $b_3 = -3$.
On the other hand, the hypercharge gauge coupling $\alpha_Y = \frac{5}{3} (\alpha/\cos^2(\theta)_W)$ evolves in general to a different value $\alpha^{'}_1 \equiv \alpha_Y(M_{32})$\,:
\begin{equation}
\frac{1}{\alpha_Y}-\frac{1}{\alpha^{'}_1} = \frac{b_Y}{2\pi}\,
{\rm \ln}\frac{M_{32}}{M_Z}\ ,
\label{eq:m32yrun}
\end{equation}

\noindent
with $b_Y = \frac{33}{5}$.
Notice that Eqs.\,(\ref{eq:m32run}) and\,(\ref{eq:m32yrun}), as are valid in ${SU(5) \times U(1)}$, are replacing Eq.\,(\ref{eq:running}), which is valid in $SU(5)$.
Above the scale $M_{32}$, the gauge group is $SU(5) \times U(1)$, with the $U(1)$ gauge couplings $\alpha_1$ related {\it across}\footnote{In contrast to Eqs.\,(\ref{eq:m32run}, \ref{eq:m32yrun}), which apply additionally for the ``run'' values $M_Z \rightarrow Q$ and $\alpha_{\{Y,2,3\}}(M_Z) \rightarrow \alpha_{\{Y,2,3\}}(Q)$, this expression is valid only at the scale $M_{32}$, encoding the discontinuity in ``$\alpha_1$'' which arises due to the re-mixing with the $U(1)$ factor emergent out of broken $SU(5)$.} $M_{32}$ to $\alpha^{'}_1 $  and the $SU(5)$ gauge coupling $\alpha_5$ by 
\begin{equation}
\frac{25}{\alpha^{'}_1} = \frac{1}{\alpha_5} + \frac{24}{\alpha_1}\ .
\label{eq:a1disc}
\end{equation}

\noindent
The $SU(5)$ and $U(1)$ gauge couplings continue to evolve above the scale $M_{32}$, eventually becoming equal at higher scale $M_{51}~(\approx\!M_U)$.
The consistency condition that ${M_{51} \ge M_{32}}$ implies $\alpha^{'}_1 \le \alpha_5(M_{32})$.
The {\it maximum possible} value of $M_{32}$ is obtained when $\alpha^{'}_1 = \alpha_5(M_{32})$ and is given by
\begin{equation}
\frac{1}{\alpha_Y} - \frac{1}{\alpha_5} = \frac{b_Y}{2 \pi} {\rm ln} \frac{M^{\rm max}_{32}}{M_Z}\ ,
\label{eq:m32max}
\end{equation}

\noindent
which is identical to the third equation of set\,(\ref{eq:running}), {\it i.e.}~$M^{\rm max}_{32} = M_{SU(5)}$.
A schematic demonstration of the prominent features of flipped gauge coupling unification is shown in Figure~1.
\vspace*{2mm}
\begin{figure}[htp]
\begin{center}
\includegraphics[width=.8\textwidth,angle=0]{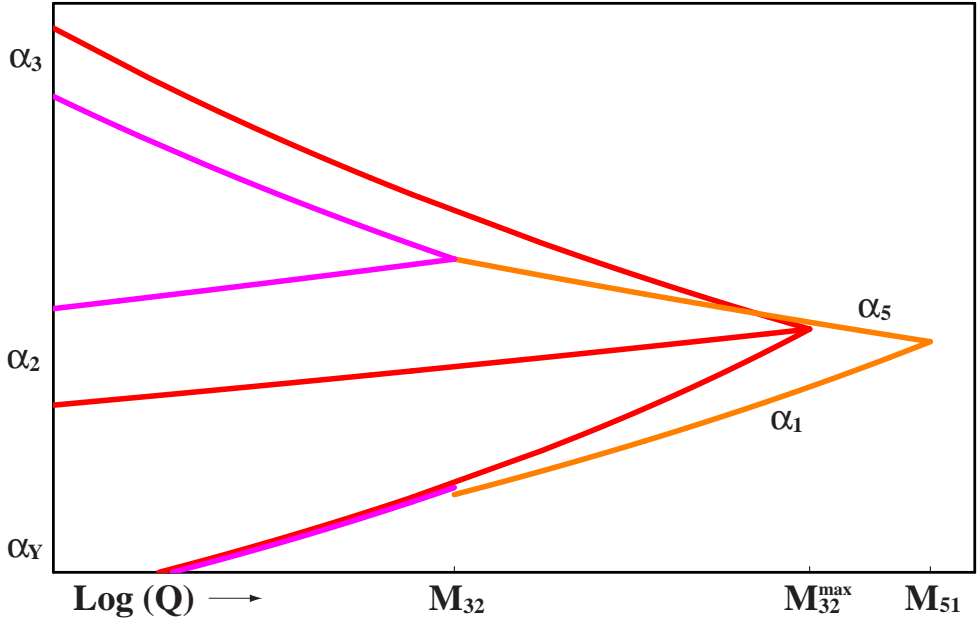}
\end{center}
\vspace{-4mm}
\caption[]{\it{\small
An exaggerated heuristic demonstration of the prominent features of flipped coupling unification.
Notice the discontinuity in the (purple) line of $U(1)_Y$, as it remixes between the ``grand unified'' $U(1)$, and that which emerges out of broken $SU(5)$ at the scale $M_{32}$.
Proceeding upward from this interim stage in orange, $SU(5) \times U(1)$ is itself unified at some higher scale $M_{51}$.
For comparison, the standard $SU(5)$ scenario is shown in red with a single unification at $M^{\rm max}_{32} \geq M_{32}$, and predicting a larger value for $\alpha_s(M_Z)$.}}
\label{fig:1}
\end{figure}
\vspace*{2mm}

Solving the above equations\,(\ref{eq:m32run},\ref{eq:m32max}) for the value of $\alpha_s(M_Z)$, we obtain~[31,32]\,:
\begin{equation}
\alpha_s(M_Z) = \frac{\frac{7}{3}\alpha_{em}}{5~\sin^2\theta_W - 1 + \frac{11}{2\pi} \alpha_{em}~{\rm ln} (\frac{M^{\rm max}_{32}}{M_{32}})}\ ,
\label{eq:asofssqt}
\end{equation}

\noindent
and since $M_{32} \leq M^{\rm max}_{32}$, we also automatically get that, for a given value of $\sin^2\theta_W$,
\begin{equation}
\alpha^{{\rm flipped}~SU(5)}_s(M_Z) \leq \alpha^{SU(5)}_s(M_Z)\ .
\label{eq:asltmax}
\end{equation}

\noindent
This result emerges as a {\it direct} consequence of the ``cohabitation'' property of the photon, as it offers the possibility of decoupling {\it somewhat} the scales at which the Standard Model $\{SU(3),SU(2)\}$ and then the $U(1)$ are unified, thus naturally allowing the strength of the $U(1)$ gauge to become smaller than in minimal $SU\!SY~SU(5)$ for the same values of $\alpha_s(M_Z)$.
This is {\it exactly} what is experimentally needed, because in minimal $SU(5)$ one is overshooting the correct value of $\alpha_s(M_Z)$ once the experimentally determined value of $\sin^2\theta_W$ is used, and the next-to-leading order corrections to Eq.\,(\ref{eq:asofssqt}) are applied by the substitution~[33]
\begin{equation}
\sin^2\theta_W \to \sin^2\theta_W - \delta_{2\!-\!{\rm loop}} - \delta_{\rm light} - \delta_{\rm heavy}\ .
\label{eq:ssqteff}
\end{equation}

\noindent
In the previous expression, $\delta_{2\!-\!{\rm loop}} \approx 0.0030$ accounts for the two-loop contributions to the renormalization group equations ($RGE$s), while $\delta_{\rm light}$ and $\delta_{\rm heavy}$ include the effects of light $SU\!SY$ thresholds and $GUT$ scale thresholds, both which can in principle carry either sign.
If one neglects $\delta_{\rm light}$ and $\delta_{\rm heavy}$, then the conventional $SU(5)$ prediction seen from Eq.\,(\ref{eq:asofssqt}) with $M_{32} = M^{\rm max}_{32}$, balloons to $\alpha_s(M_Z) \approx 0.130$, about $6\!-\!\sigma$ away from the experimentally determined value given in Eq.\,(\ref{eq:consts})!
Furthermore, $\delta_{\rm light} > 0$ in large regions of parameter space, and $d\!=5\!$ proton decay constraints given in Eq.\,(\ref{eq:kplusnulim}) entail $\delta_{\rm heavy} > 0$, which can only make the situation much worse, as emphasized previously~[21].
In the case of flipped $SU(5)$ things are much different.
Let us first calculate a reasonable range for $M_{32}$~[32], using Eq.\,(\ref{eq:asofssqt}) with $\sin^2\theta_W$ replaced by $\sin^2\theta_W - \delta_{2\!-\!{\rm loop}}$, leaving for later the inclusion of $\delta_{\rm (light,heavy)}$.
\vspace*{2mm}
\begin{figure}[htp]
\begin{center}
\includegraphics[width=.8\textwidth,angle=0]{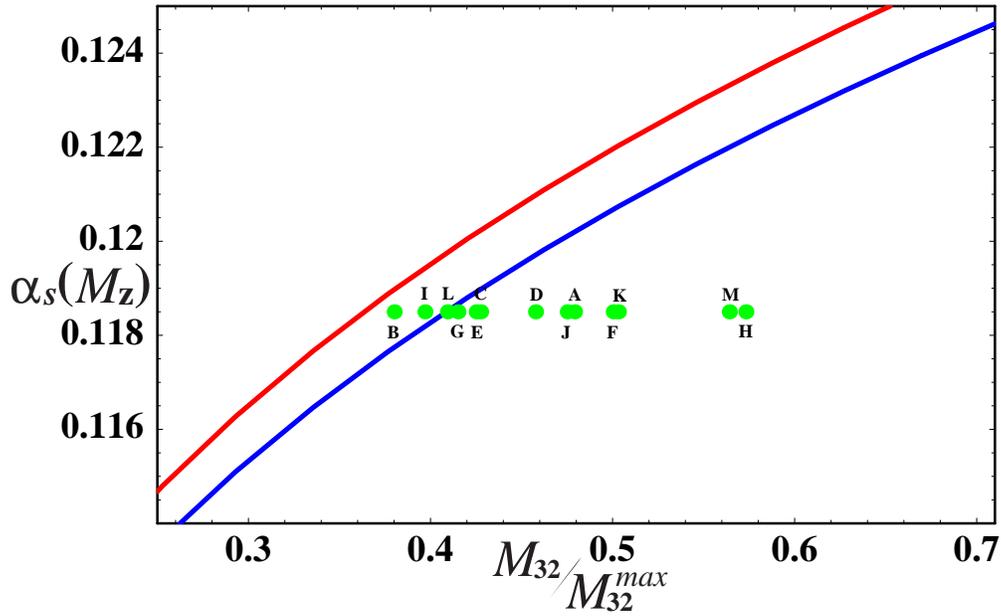}
\end{center}
\vspace{-4mm}
\caption[]{\it{\small
The solid lines show the flipped $SU(5)$ correlation between $M_{32}$
and $\alpha_s(M_Z)$, taking $\sin^2\theta_W^{\overline{MS}} = 0.23117 \pm 0.00016$,
and including $\delta_{\rm 2loop}$.
The shifted points represent the effects of $\delta_{\rm light}$ calculated at the $C\!M\!S\!S\!M$ benchmarks for central values of $\alpha_s(M_Z)$ and $\sin^2\theta_W$.}}
\label{fig:2}
\end{figure}
\vspace*{2mm}

\noindent
Figure~2~[32] exhibits the correlation between $M_{32}$ and $\alpha_s(M_Z)$ in flipped $SU(5)$.
The solid lines indicate the range of values of $M_{32}$ allowed for a given value of $\alpha_s(M_Z)$ (as given in the ${\overline{MS}}$ prescription), assuming the experimentally allowed range of $\sin^2\theta^{\overline{MS}}_W$ as given in Eq.\,(\ref{eq:consts}), and setting $\delta_{\rm (light,heavy)} = 0$.
Taking the central experimental values of $\alpha_s(M_Z)$ and $\sin^2\theta_W$, again from Eq.\,(\ref{eq:consts}), we see immediately that $M_{32}$ must be {\it significantly} lower than it's maximum standard $SU(5)$ value of $M^{\rm max}_{32} \approx 20.3 \times 10^{15}~GeV$.
Next, let us explore the possible consequences of $\delta_{\rm light}$ for $M_{32}$.
The $\delta_{\rm light}$ correction may be approximated by~[33,31,32]\,:
\begin{eqnarray}
\delta_{\rm light}&=&{\alpha\over20\pi}\Bigl[
-3L(m_t)+{28 \over 3}L(m_{\tilde g})-{32 \over 3}L(m_{\tilde w})
-L(m_h)-4L(m_H)\qquad
\nonumber\\
&&+{5 \over 2}L(m_{\tilde q})-3L(m_{\tilde \ell_L})
+2L(m_{\tilde \ell_R})-{35 \over 36}L(m_{\tilde t_2})
-{19 \over 36}L(m_{\tilde t_1})\Bigr]\ ,\qquad
\label{eq:dlight}
\end{eqnarray}

\noindent
where $L(x) = \textrm{ln}(x/M_Z)$.
Clearly, this effect is highly dependent on the detailed nature of sparticle spectrum.
As usual, we assume that the soft supersymmetry-breaking scalar masses $m_0$, gaugino masses $m_{1/2}$ and trilinear coefficents $A_0$ are universal at the $GUT$ scale, ({\it i.e.}~the {\it Constrained} $M\!S\!S\!M$).
One then is able to calculate the sparticle spectra in terms of these quantities and in addition, $\tan \beta \equiv \langle \bar{h}\rangle/\langle h\rangle$ (see Eq.\,(\ref{eq:yuksup})), and the sign of $\mu$ (see Eq.\,(\ref{eq:gutsup})), with $m_{t} = 175~GeV$, as is done for example in references~[34,35].
Before making a more general survey, first recall that a number of {\it benchmark} $C\!M\!S\!S\!M$ scenarios have been proposed~[35], which include theoretical constraints, {\it e.g.}~electroweak symmetry breaking triggered by radiative corrections~[15], and are consistent with all the experimental limits on sparticle masses as well as the $LEP$ lower limit on $m_h$, the world-average value of $b \rightarrow s~\gamma$ decay, the preferred range $0.1 < \Omega_x~h^2 < 0.3$ of the supersymmetric relic density\footnote{This relic density is composed of the Lightest Supersymmetric Particle ($LSP$), which needs to be the neutralino, ($\chi^0$)~[18].}, and a value for the anomalous magnetic moment of the muon, $g_{\mu}\!-\!2$, within $2\!-\!\sigma$ of the present experiments.
These points all have $A_0 = 0$, but otherwise span the possible ranges of $m_0$, $m_{1/2}$ and $\tan \beta$, and feature both signs for $\mu$.
Figure~2 also shows the change in $M_{32}$ induced by the values of $\delta_{\rm light}$ within these benchmark models, assuming a fixed value $\alpha_s(M_Z) = 0.1185$, as indicated in Eq.\,(\ref{eq:consts}).
In general, these bechmark models {\it increase} $M_{32}$ for any fixed value of $\alpha_s(M_Z)$ and $\sin^2\theta_W$, as the sparticle masses {\it increase}~[32].
In other words, the {\it lighter} the sparticle spectra is, ({\it e.g.}~benchmark models {\it B, I, L,} $\ldots$~[35]), the {\it smaller} the needed value of $M_{32}$ is.
Of course, the {\it smaller} $M_{32}$ becomes, the {\it shorter} the proton lifetime gets, and this can make things very interesting experimentally.
The reason for this is that while $d\!=\!5$ proton decay, $p \rightarrow K^+\nu$, is banned from flipped $SU(5)$ thanks to it's economic missing partner mechanism as discussed above, the $d\!=\!6$ proton decay, $p \rightarrow (e^+/\mu^+)\,\pi^0$, is still on~[30,31,32].
Indeed, one finds in this case~[36,21]\,:
\begin{equation}
\tau~(p \to e^+ \pi^0) \,=\,
3.8 \times 10^{35} \left( {M_{32} \over 10^{16}~{\rm GeV}} \right)^4
\left( {\alpha_5(M^{\rm max}_{32}) \over \alpha_5(M_{32})} \right)^2
\left( {0.015~{\rm GeV}^3 \over
\alpha} \right)^2~{\rm y}\ ,
\label{eq:taupepi}
\end{equation}

\noindent
where reference values for $M_{32}$ and $\alpha_5(M_{32})$, as well as the relevant matrix elements $\alpha$ and $\beta$ have been absorbed, and
\begin{equation}
\frac{\alpha_5(M^{\rm max}_{32})}{\alpha_5(M_{32})} = 1 - \frac{33~\alpha_5(M^{\rm max}_{32})}{28 \cdot 2 \pi}~{\rm ln}(M_{32}/M^{\rm max}_{32})\ .
\label{eq:a5ratio}
\end{equation}

Concerning the proton decay modes, some characteristic flipped $SU(5)$ predictions can be made~[30,36,32]\,:
\begin{eqnarray}
\Gamma(p \rightarrow e^+ \pi^o) \;=\; \frac{\cos ^2 \theta_c}{2}
 |U_{\ell_{11}}|^2
\Gamma(p \rightarrow {\bar \nu} \pi^+) \;=\; \cos ^2 \theta_c
|U_{\ell_{11}}|^2
\Gamma(n \rightarrow {\bar \nu} \pi^o)
\nonumber\\
\Gamma(n \rightarrow e^+ \pi^-) \;=\; 2
 \Gamma(p \rightarrow e^+ \pi^o) \quad;\quad
\Gamma(n \rightarrow \mu^+ \pi^-) \;=\; 2
\Gamma(p \rightarrow \mu^+ \pi^o)
\\
\Gamma(p \rightarrow \mu^+ \pi^o) \;=\; \frac{\cos ^2 \theta_c}{2}
 |U_{\ell_{12}}|^2
\Gamma(p \rightarrow {\bar \nu} \pi^+) \;=\; \cos ^2 \theta_c
|U_{\ell_{12}}|^2
\Gamma(n \rightarrow {\bar \nu} \pi^0)\ ,
\nonumber
\label{eq:gammaepi}
\end{eqnarray}

\renewcommand{\baselinestretch}{1.04}
\noindent
where $U_{\ell}$ is the charged lepton mixing matrix that ``turns'' mass eigenstates ($\ell_F$) into leptonic flavor eigenstates ($\ell_L$), as: $\ell_L = \ell_F~U_{\ell}$~[36].
In the light of recent experimental evidence for near-maximal neutrino-mixing~[12], it is reasonable to think that (at least some of) the ($e^+/\mu^+$) entries in $U_{\ell}$ are $O(1)$, and thus it looks reasonable to expect~[32] very similar decay rates for both processes $p \rightarrow (e^+/\mu^+)\,\pi^0$, at the level of a few times $10^{35}$ years, as long as the sparticle mass spectrum is not on the heavy side~[32].
But then, how do we know that the sparticle spectrum is not heavy?
Here, the recent developments in $g_{\mu}\!-\!2$, both in theory and experiment, when applied to flipped $SU(5)$, come to our rescue.
Using the new $BNL\!-\!E821$ data~[37], as well as a consensus value for the theoretical estimate of $\alpha_{\mu} \equiv \frac{g_{\mu}\!-\!2}{2}$ in the Standard Model based on low-energy $e^+e^-$ data, one gets~[38] $\delta_{\alpha_{\mu^+}} \equiv (\alpha_{\mu^+})_{exp} - (\alpha_{\mu^+})_{SM} = +~33.9~(11.2) \times 10^{-10}$, a $3\!-\!\sigma$ effect!
Such an effect can be easily accomodated within the $M\!S\!S\!M$, pointing to a {\it light} $SU\!SY$ spectrum, as is given by a loop-effect, in conjunction with large $\tan \beta$ and a positive sign for $\mu$~[39].
About ten years ago we noticed~[39] that in the context of flipped $SU(5)$ one might expect to observe significant deviations of $g_{\mu}\!-\!2$ from the Standard Model within the then under consideration $BNL\!-\!E821$ experiment, for the regime of light sparticles, large $\tan \beta$ and $\mu > 0$!
In sharp contrast, we also noticed~[39] that for minimal $SU\!SY~SU(5)$, $SO(10)$, $\ldots$ , one should not expect large contributions to $\alpha_{\mu}^{SU\!SY}$, since the constraint from $d\!=\!5$ proton decay required heavy sparticles, in this case sleptons, as well as $\tan \beta \leq 5$.
In other words, even post-$SU(5)$ models that strive to tame the menace of $d\!=\!5$ proton decay are coming {\it very short} in terms of contributions to $g_{\mu}\!-\!2$.
\vspace*{2mm}
\begin{figure}[ht]
\begin{center}
\includegraphics[width=.8\textwidth,angle=0]{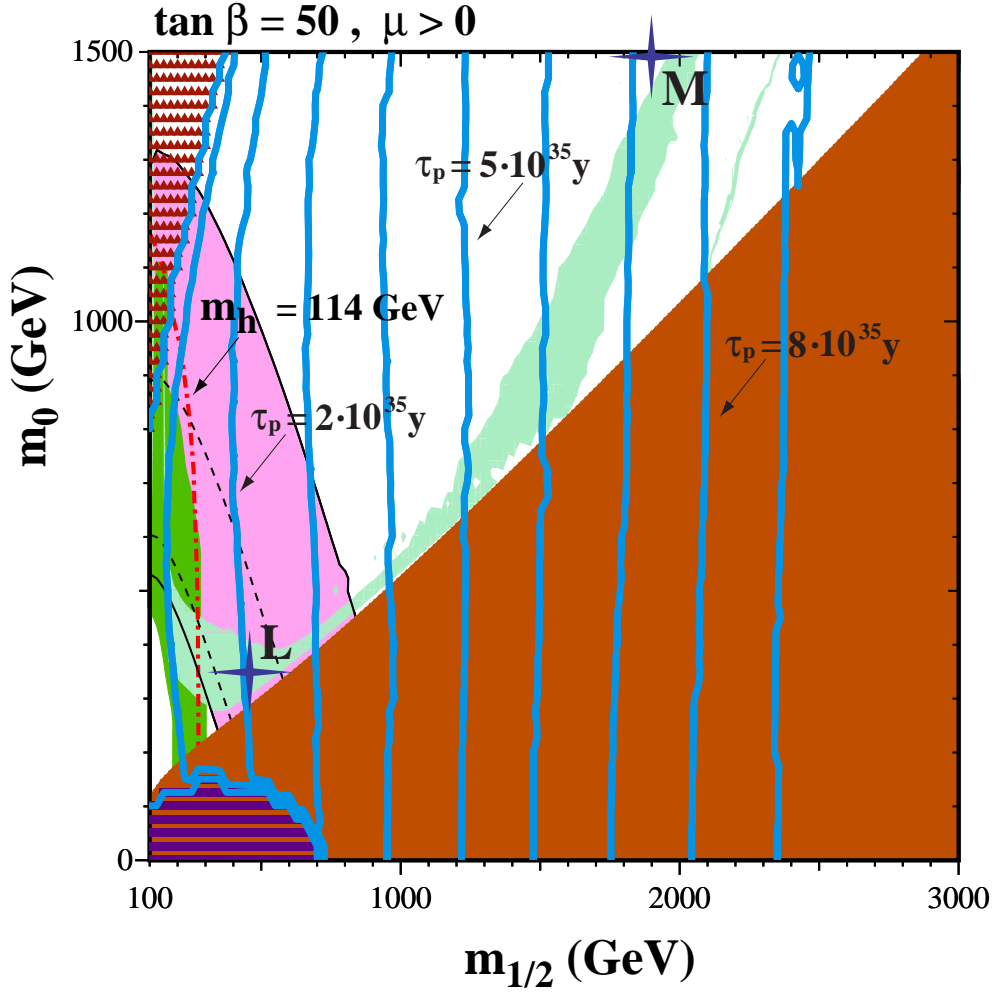}
\end{center}
\vspace{-4mm}
\caption[]{\it{\small
The light blue lines trace contours of $\tau~(p\!\rightarrow\!(e^+/\mu^+)\,\pi^0)$ as it varies with the value of $\delta_{\rm light}$ scanned across the $(m_{1/2},m_0)$ plane.}}
\label{fig:3}
\end{figure}
\vspace*{2mm}

A typical example of flipped $SU(5)$ proton decay in the $C\!M\!S\!S\!M$ is presented in Figure~3.
The solid light blue lines are contours of $\tau~(p\!\rightarrow\!(e^+/\mu^+)\,\pi^0)$ in the ($m_{1/2}, m_0$) plane for the $C\!M\!S\!S\!M$ with $\tan \beta = 50$ and positive $\mu$.
The dark blue cross ($L$) locates the corresponding $C\!M\!S\!S\!M$ benchmark point~[35], which also bears $\tan \beta = 50$ and $\mu > 0$, within the plane.
The large lower-right area shaded in dark orange is excluded for the presence of a charged $LSP$\footnote{By definition, an electrically charged $LSP$ would scatter light, and could not then constitute ``dark'' matter.}, the small green shaded region at low $m_{1/2}$ is excluded by $b \rightarrow s~\gamma$, electroweak symmetry breaking is not possible in the upper-left hatched red area, and the blue horizontally-striped region at low $m_0$ has tachyons.
The light turquoise band marks the $0.1 < \Omega_x~h^2 < 0.3$ preferred region, while the pink vertical stripe bounded by dashed ($-\!-\!-$) lines is consistent with $g_{\mu}\!-\!2$ at $1\!-\!\sigma$, and the outer solid lines denote the $2\!-\!\sigma$ limits.
The dot-dashed ($-\!\cdot\!-\!\cdot\!-$) red line corresponds the $LEP$ lower limit on the Higgs mass, of $m_h = 114~GeV$.
We can take from Figure~3 that the ``bulk'' regions of the parameter space preferred by astrophysics and cosmology, and by the presently accepted value of $g_{\mu}\!-\!2$~[37,38], which occur at relatively small values of ($m_{1/2}, m_0$), generally correspond to $\tau~(p\!\rightarrow\!(e^+/\mu^+)\,\pi^0) \sim (1\!-\!2) \times 10^{35}~y$\footnote{Again, it is emphasized that this estimate, as well as the general variations of $\tau_p$ witnessed in the ($m_{1/2}, m_0$) plane, are derived out of the accompanying $\delta_{\rm light}$ threshold corrections, to the exclusion of those from $\delta_{\rm heavy}$.}.
Interestingly enough, such values for the proton lifetime, as per this decay mode, are within the reachable range of a new generation of massive Megatonnes water-{\v C}erenkov detectors that are being proposed~[40]!
Let me mention that if we use as a $L$amp-post the $L$-benchmark point of Figure~3, we get $\Omega_xh^2 \approx 0.21$, $\delta_{\alpha_{\mu}} = +~31 \times 10^{-10}$, $m_h \approx 118~GeV$, $m_{(LSP~\equiv~\chi^0)} \approx 188~GeV$, $m_{\tilde{\tau}_1} \approx 242~GeV$, $m_{\tilde{t}_1} \approx 714~GeV$, $m_{\tilde{g}} \approx 994~GeV$, {\it etc.}, {\it etc.}
This is a rather light sparticle spectrum, observable in present or near future accelerators.
It is highly remarkable that, as clearly depicted in Figures~(2,3), for the case of flipped $SU(5)$, there are very strong correlations between physics at the $GUT$ scale and physics at the Electroweak scale.
The sparticle mass spectrum, through the $\delta_{\rm light}$ contributions to the value of $\sin^2\theta_W$ in Eq.\,(\ref{eq:asltmax}), participates in the evaluation of $\alpha_s(M_Z)$ in Eq.\,(\ref{eq:asofssqt}), and thus in the determination of $M_{32}$, as shown in Figure~2.
The value of $M_{32}$ fixes in turn {\it both} the proton lifetime through Eq.\,(\ref{eq:taupepi}), as shown in Figure~3, {\it and} the neutrino mass spectrum, through Eq.\,(\ref{eq:twostep}), which then via $U_l$ in Eq.\,(\ref{eq:gammaepi}), determines the proton decay branching ratios!
A {\it light} sparticle spectrum as potentially suggested by the present value of $g_{\mu}\!-\!2$~[37,38], entails a ratio $M_{32}/M_{32}^{max}$ much smaller than unity, and thus perhaps an {\it observably short} proton lifetime, as well as a rather {\it light} neutrino mass spectrum from Eq.\,(\ref{eq:twostep}), avoiding contradictions with experiment~[12].
The case of $GUT$ thresholds can be treated similarly, as in~[31,32].
Here it suffices to notice that $\delta_{\rm heavy}$ leads to a simple rescaling of the $M_{32}/M_{32}^{max}$ factor\,:
\begin{equation}
{M_{32}\over M^{\rm max}_{32}}\to {M_{32}\over M^{\rm max}_{32}}
\ e^{-10\pi\,\delta_{\rm heavy}/11\alpha}\ .
\label{eq:heavyshift}
\end{equation}

\noindent
Assuming the plausible ranges $-0.0016 < \delta_{\rm heavy} < .0005$, one sees that $M_{32}/M_{32}^{max}$ is rescaled in the range ($0.8$ to $1.8$), thus leading to an overall proton lifetime ($\tau_p$) rescaling in the range ($.4$ to $10$).
Until now I have followed a ``bottom-up'' approach, and have tried hard to build a favorable Flipped Phenomenological, or {\it F-enomenological}, case.
For that reason, I have avoided dealing with a rather obvious drawback of $SU(5) \times U(1)$: it is not Grand Unified!!!
It instead contains the extra factor of $U(1)$ which spoils the very definition of a $GUT$!
In addition, as Eq.\,(\ref{eq:yuksup}) indicates, we lose the hard-won relation of Eq.\,(\ref{eq:lambdatau})~[3,4,5], to be replaced only by Eq.\,(\ref{eq:nudiraca}), which implies equal Dirac-type neutrino and up-type quark masses.
{\it So, what's going on}?

\section{\sectionfont F-INALE}

Well, at this point I have to bring in the ``top-down'' approach.
It is rather well known  that the modern and more articulate form~[25] of flipped $SU(5)$ arises out of heterotic string theory in the Free-Fermionic Formulation.
The key point here is that string theory at the perturbative $k\!=\!1$ level permits no adjoint or larger representations~[23], as would be required to hold the Higgs multiplets within the ``usual $GUT$ suspects''.
This means that $SU(5)$, $SO(10)$, $E_6$, $\ldots$ will not fit, and thus we must acquit!
It is only $SU(5) \times U(1)$, in it's close resemblance to the lower-energy success of $SU(2)_L \times U(1)_Y$, that needs no adjoints or other large representations, but only those in which the quarks and leptons reside\footnote{To put this simply, when an element of the Higgs multiplets take on a non-zero expectation value, this breaks the symmetries associated with any quantum numbers carried in a non-scalar sense.  Since the flipped assignments of Eq.\,(\ref{eq:flippedcontent}) carry neutrinos in {\it both} the $\mathbf{\bar{5}}$ and $\mathbf{10}$, they are {\it both} acceptable as Higgs representations, in contrast to Eq.\,(\ref{eq:smcontent}).}.
Thus, $SU(5) \times U(1)$ was taken seriously and ``string derived'' versions were soon developed~[41,42].
However, as with any particle model coming from the string, flipped $SU(5)$ must be sitting within a larger group, such as $E_8 \times E_8$ in $D=10$, or $SO(44)$ in the direct $D=4$ Free-Fermionic construction.
As such, at some stage $SU(5) \times U(1)$ existed within a {\it virtual} $SO(10)$, thus retaining some ``memory'' of this origin, and picking up {\it for free} all the good properties of ``true'' $GUT$s which were exposed earlier (charge quantization, $\lambda_b = \lambda_{\tau} \vert_M$, {\it etc.}), and so rendering a respectable mass spectra~[43] without the need for adjoint or larger Higgs representations for it's breaking down to the Minimal Supersymmetric Standard Model.
In actuality, {\it charge quantization} is not guaranteed in superstring models, since many of them are known to predict the existence of particles with charges $\frac{1}{n}$, where $n$ is some model dependent integer~[44].
One can show~[45] in fact, that their presence is a {\it generic} feature of models derived from the superstring with a level $k\!=\!1$, {\it i.e.}~realistic models.
Fortunately, there is another way to avoid {\it detectable} fractional charged particles, namely to confine them analogously to quarks in QCD ($SU(3)_c$)!
It is highly remarkable that in string derived flipped $SU(5)$~[41], we have proven~[41,46,47] that there are {\it no} free fractional states at {\it any} mass level, as they are all {\it confined} by an accompanying ``hidden'' $SU(4)$ gauge group which becomes strong at some large mass scale $\Lambda_4~(\sim 10^{12}\!-\!10^{13}~GeV)$.
On the other hand, the ``hidden'' sector of the flipped $SU(5)$ string model {\it does} contain integer-charged bound states called ``Cryptons''~[47], after the manner of protons, which are metastable and {\it may} provide candidates for Dark Matter or the long-sought source for Ultra High Energy Cosmic Rays~[48,49,50].
Such a fractional charge confinement, {\it \`{a} la} QCD, is so far a unique characteristic of flipped $SU(5)$, and has not been realized in other realistic string derived scenarios, {\it e.g.}~the Standard Model~[51].
So, in order for {\it charge quantization} to remain kosher, string theory again points directly to flipped $SU(5)$.

More recently, in the modern framework of non-pertubative String\,/$M$-theory, $SU(5) \times U(1)$ has once more naturally appeared~[52] via {\it virtual} $SO(10)$ in a Horava-Witten type description~[53], as well as through stacks (two or three) of intersecting $D\!-\!6$ branes which yield a reasonable, though enlarged, fermionic particle spectra~[54].
Flipped $SU(5)$ has also appeared in {\it orbifolded} $D\!=\!5$ $SO(10)$~[55], $SU(7)$ and $SO(14)$, where in the case of the last two groups, unification of families is also achieved~[56].

The message that comes clearly from the above analysis is that $SU(5) \times U(1)$ should be considered, at best, as an {\it effective} theory below the $GUT$ scale, exactly as $SU(2)_L \times U(1)_Y$ is considered as an {\it effective} theory below the Electroweak scale (see Table~2).
All in all, the several strong observable correlations between otherwise unrelated low energy phenomena mentioned above may make or break flipped $SU(5)$.
{\it La Lutte Continue} $\ldots$

\section*{\sectionfont Acknowledgements}
I would like to thank Joel Walker and Eric Mayes for a critical reading of the manuscript and for their comments.
This work was partially supported by DOE grant DE-F-G03-95-ER-40917.

\end{fmffile}

\end{document}